\documentclass[numberedappendix]{emulateapj}
\usepackage{amssymb,amsmath,amsthm}
\usepackage{longtable}
\usepackage{amssymb}
\usepackage{amsmath, xspace}
\usepackage{graphics,graphicx} 
\usepackage{rotating}
\usepackage{color}
\usepackage{perpage} 
\renewcommand{\thefootnote}{\arabic{footnote}}
\renewcommand{\thefootnote}{\arabic{footnote}}
\usepackage{url}
\usepackage[colorlinks=true,citecolor=blue ]{hyperref}

\begin{document}


\title{Searching for the 3.5 keV Line in the Stacked Suzaku Observations of Galaxy Clusters}


\author{
Esra~Bulbul\altaffilmark{1},
Maxim~Markevitch\altaffilmark{2},
Adam~Foster\altaffilmark{3},
Eric~Miller\altaffilmark{1},
Mark~Bautz\altaffilmark{1},
Mike~Loewenstein\altaffilmark{2},
Scott~W.~Randall\altaffilmark{3}, and
Randall~K.~Smith\altaffilmark{3}
}
\altaffiltext{1}{Kavli Institute for Astrophysics \& Space Research, Massachusetts Institute of Technology, 77 Massachusetts Ave, Cambridge, MA 02139, USA\\}
\altaffiltext{2}{NASA Goddard Space Flight Center, Greenbelt, MD, USA\\}
\altaffiltext{3}{Harvard-Smithsonian Center for Astrophysics, 60 Garden Street, Cambridge, MA~02138, USA\\}

\email{Email: ebulbul@mit.edu}

\begin{abstract}

We perform a detailed study of the stacked {\it Suzaku} observations of 47 galaxy clusters, spanning a redshift range of 0.01--0.45, to search for the unidentified 3.5 keV line. This sample provides an independent test for the previously detected line. We detect a $2\sigma$-significant spectral feature at 3.5 keV in the spectrum of the full sample. When the sample is divided into two subsamples (cool-core and non-cool core clusters), cool-core subsample shows no statistically significant positive residuals at the line energy. A very weak ($\sim2\sigma$ confidence) spectral feature at 3.5 keV is permitted by the data from the non-cool core clusters sample. The upper limit on a neutrino decay mixing angle of $\sin^{2}(2\theta)=6.1\times10^{-11}$ from the full {\it Suzaku} sample is consistent with the previous detections in the stacked {\it XMM-Newton} sample of galaxy clusters (which had a higher statistical sensitivity to faint lines), M31, and Galactic Center at a 90\% confidence level. However, the constraint from the present sample, which does not include the Perseus cluster, is in tension with previously reported line flux observed in the core of the Perseus cluster with {\it XMM-Newton} and {\it Suzaku}.

\end{abstract}

\section{Introduction}
\label{sec:intro}

The detection of an unidentified emission line near 3.5 keV in the stacked {\it XMM-Newton} observations of galaxy clusters and in the Perseus cluster has received significant attention from astrophysics and particle physics communities \citep[][Bu14a hereafter]{b14}. The detection was also reported in the outskirts of the Perseus cluster and the Andromeda galaxy observed with {\it XMM-Newton}  \citep[][Bo14 hereafter]{bo14}, as well as  in {\it Suzaku} observations of the Perseus cluster core \citep[][see, however, a nondetection by Tamura et al.\ 2015]{urban15, franse16}. An emission line at a consistent energy was detected in the {\it XMM-Newton} and {\it Chandra} observations of the Galactic center and in eight other individual clusters  \citep{bo15,jp15,iakubovskyi15}.

Although the line was detected by several X-ray detectors in a variety of objects, the origin of the line is unclear. Bu14a has discussed potential astrophysical origin of this line, e.g., an emission line from the nearby weak atomic transitions of  K {\sc xviii} and Ar {\sc xvii} dielectronic recombination (DR); they found that these lines have to be 10--20 times above the model prediction. \citet{jp15} and \citet{carlson15} suggested that a large fraction of cool gas with $T<1$ keV in cluster cores may produce lines from K {\sc xviii} stronger then those Bu14a allowed for. We commented in \citep[][hereafter Bu14b]{b14b} that ratios of the observed lines from other elements exclude significant quantities of such cool gas. Recently, \citet{gu15} suggested that charge exchange between Sulfur ions and neutral gas, a process not included in Bu14a, may produce excess near 3.5 keV. These as well as some other recent spatially-resolved studies are reviewed by \citet{franse16}.

A more exotic possibility that is interesting to consider is that the observed line is a signal from decaying dark matter particles \citep{abazajian14,horiuchi16}. In previous studies, they reported that the flux of the line is consistent across objects of different mass (Andromeda galaxy, stacked galaxy clusters, and Galactic center) when the mass scaling in decaying dark matter models are taken into account \citep[see][]{bo15}. Although it is challenging to test this hypothesis with the current CCD (100--120 eV) resolution X-ray telescopes, the radial distribution of the line in a well exposed galaxy cluster may provide further information on its origin. \citet{franse16} examined the flux distribution of the 3.5 keV line as a function of radius in the Perseus cluster. However, the observed line flux from the Perseus core ($r\leq1^{\prime}$) appears to be in tension with the detection from other objects, assuming the decaying dark matter model (Bu14a, Franse et al.\ 2016). \citet{franse16} found that the profile of the line is consistent with a dark matter origin as well as with an unknown astrophysical line. Recently, \citet{ru15} have analyzed a very deep {\it XMM-Newton} observation of the Draco dwarf galaxy. They found no line signal in the spectrum from the MOS detectors and a $2.3\sigma$-significant hint of a positive signal at the right energy in the independent PN spectrum, both findings consistent with the previous detections within uncertainties.

Bu14a have laid the framework for stacking X-ray observations at the rest frame and successfully applied their method to a large sample of {\it XMM-Newton} observations. In this work, we take a step further to search for the unidentified line in the stacked {\it Suzaku} observations of 47 galaxy clusters. This paper is organized as follows. Section 2 describes the data processing, and spectra stacking. In section 3 and 4 we provide our results and conclusions. All errors quoted throughout the paper correspond to 68\%(90\%) single-parameter confidence intervals; upper limits are at 90\% confidence, unless otherwise stated. Throughout our analysis we used a standard $\Lambda$CDM cosmology with $H_{0}$ =71 km s$^{-1}$ Mpc$^{-1}$, $\Omega_{M}= 0.27$, and $\Omega_{\Lambda}= 0.73$. In this cosmology, 1$^{\prime}$ corresponds to $\sim$0.11 Mpc at redshift of 0.1.

\section{Sample Selection and Data Reduction}
\label{sec:analysis}

In an attempt to smooth the instrumental and background features related to the {\it Suzaku} XIS detectors, we select a sample of galaxy clusters based on the number of X-ray counts in their 2--10 keV band. To be able to smear the instrumental features by blue-shifting the spectra to the source frame, we select clusters covering a large redshift range of $0.01<z<0.45$. A significant number of on-axis X-ray observations of galaxy clusters have been performed by {\it Suzaku} since the launch in 2005. We selected observations with a minimum of 10,000 counts in z $<$ 0.2 per cluster, and 5,000 counts per cluster for clusters with redshifts $0.2 < z < 0.44$. The final sample includes 51 {\it Suzaku} X-ray observations of 47 galaxy clusters. The details of the observations are summarized in Table \ref{table:obs} together with the filtered exposure times. The filtering process is described below.
We note that the Perseus cluster, the X-ray brightest cluster, has the longest observations (1Ms) available in the {\it Suzaku} archive. However, to avoid the final stacked spectrum being dominated by this cluster we exclude it from our sample. The flux distribution of the 3.5 keV line out to the virial radius of the Perseus cluster has already been studied in great detail in \citet{franse16}.

The details of {\it Suzaku} data reduction are described in \citet{bu2016} and \citet{franse16}. Here, we provide a summary of the steps we follow in the data analysis. After the calibrated data is filtered from the background flares, source images in the 0.4--7.0 keV band are extracted from the filtered event files. These images are used to detect point sources within the {\it Suzaku} the field-of-view (FOV) using the CIAO's tool {\it wavdetect}. The detected point sources are excluded from the further analysis.

The source and particle background spectra are extracted from the filtered event file and filtered night-time Earth data using the FTOOL {\it xisnxbgen}. The spectra are extracted within the overdensity radius $R_{500}$\addtocounter{footnote}{-3},\footnote{The overdensity radius $R_{500}$ is defined as the radius within which the average matter density of the cluster is 
500 times the critical density of the Universe at the cluster redshift.} if the estimated $R_{500}$ falls within FOV of XIS. 

The overdensity radii ($R_{500}$) are calculated using the mass-temperature scaling relation for each cluster \citep{vikhlinin2009}. The temperatures used in these estimates are obtained from previously published results in the literature. For some of the nearby clusters $R_{500}$ is larger than the XIS FOV. For those we use the largest possible region (a circle with a radius of 8.3$^{\prime}$) that encompasses the cluster center while avoiding the detector edges. The extraction radii for the full sample are given in Table \ref{table:prop}. Redistribution matrix files (RMFs) and ancillary response files (ARFs) are constructed using the FTOOLs {\it xisarfgen} and {\it xisrmfgen}.

The particle induced background spectrum is subtracted from each source spectrum prior to fitting. 
Following the same approach presented in Bu14, we first perform the spectral fitting in the Fe K-$\alpha$ band (5.5--7.5 keV rest frame) with a single temperature thermal model ({\it apec}) to determine the best-fit redshift of each cluster with the AtomDB version 2.0.2 \citep{smith01, foster12}. XSPEC v12.9.0 is used to perform the spectral fits \citep{arnaud96} with the extended $\chi^{2}$ statistics as an estimator of the goodness-of-fits. The spectral counts in each energy bin were sufficiently high to allow the use of the Gaussian statistics in this analysis \citep{protassov02}.

We combine front illuminated (FI) XIS0 and XIS3 data to increase the signal-to-noise, while the back illuminated (BI) XIS1 data are modeled independently due to the difference in energy responses. The best-fit redshifts ($z_{best}$) obtained from FI observations are given in Table \ref{table:obs}. The best-fit redshifts measured from BI observations are in good agreement with FI observations.

In order to detect a weak spectral feature such as the $\sim$3.5 keV line ($\sim1\%$ excess over the continuum), the detector and background artifacts must be eliminated from the high signal-to-noise stacked galaxy cluster spectrum. In order to accomplish this, we stacked the spectra of our selected 47 clusters at the source frame using the best-fit X-ray redshift of each observation determined above. The energies of the source and background X-ray events are rescaled to the source frame using the best-fit redshifts. The spectra within R$_{500}$ are extracted from these rescaled event and background files, before being stacked. The individual RMFs and ARFs are then remapped to the source frame. The weighting factors ($\omega_{cnts}$), given in Table \ref{table:prop}, for stacking RMFs and ARFs are calculated using the total counts in the fitting band (2--10 keV). The weighted and remapped ARFs and RMFs are combined using the FTOOLs {\it addarf} and {\it addrmf}, while {\it mathpha} is used to produce stacked source and background spectra. At the end of the stacking processes, we obtain a total of 5.4 Ms FI and 2.7 Ms BI  galaxy cluster observations in the full sample. These count-weighted response files are used in modeling the continuum and the known plasma emission lines (see Section \ref{sec:results}).

\section{Results}
\label{sec:results}

As in B14a, We fit the background-subtracted stacked source spectra with line-free multi-temperature {\it apec} models to represent the continuum emission with high accuracy. Gaussian models are added to account for individual atomic lines in the 1.95 -- 6 keV energy band. 
 Our total model includes the following lines at their rest energies: Al \textsc{xiii} (2.05 keV), Si \textsc{xiv} (2.01 keV, 2.37 keV, and 2.51 keV), Si \textsc{xii} (2.18 keV, 2.29 keV, and 2.34 keV), S \textsc{xv} (2.46 keV, 2.88 keV, 3.03 keV), S \textsc{xvi} (2.62 keV), Ar \textsc{xvii} (triplet at 3.12 keV, 3.62 keV, 3.68 keV), Cl {\sc XVI} (2.79 keV),  Cl {\sc xvii} (2.96 keV), Cl  {\sc xvii} (3.51 keV), K \textsc{xviii} (triplet 3.47 keV, 3.49 kev and 3.51 keV), K \textsc{xix} (3.71 keV), Ca \textsc{xix} (complex at 3.86 keV, 3.90 keV, 4.58 keV), Ar \textsc{xviii} (3.31 keV, 3.93 keV), Ca \textsc{xx} (4.10 keV), Cr \textsc{xxiii} (5.69 keV). 

{
{\renewcommand{\arraystretch}{1.25}
\begin{longtable*}{llllccccc}
\caption{On -Axis Galaxy Cluster Observations Performed by {\it Suzaku} \label{table:obs}}\\
\hline \hline\\
 \multicolumn{1}{l}{\textbf{Cluster}} & \multicolumn{1}{c}{\textbf{ RA }} & \multicolumn{1}{c}{\textbf{DEC}}& \multicolumn{1}{c}{\textbf{ObsID }} & \multicolumn{1}{c}{\textbf{ FI}}  & \multicolumn{1}{c}{\textbf{ BI}} & \multicolumn{1}{c}{\textbf{$z_{best}$ }} & \multicolumn{1}{c}{\textbf{ Sub-Sample }}\\ 

\multicolumn{1}{c}{\textbf{}} & \multicolumn{1}{c}{\textbf{ }} & \multicolumn{1}{c}{\textbf{}}& \multicolumn{1}{c}{\textbf{ }} & \multicolumn{1}{c}{\textbf{ Exp}} &\multicolumn{1}{c}{\textbf{Exp }} & \multicolumn{1}{c}{\textbf{}} & \multicolumn{1}{c}{\textbf{  }}\\

\multicolumn{1}{c}{\textbf{}} & \multicolumn{1}{c}{\textbf{ }} & \multicolumn{1}{c}{\textbf{}}& \multicolumn{1}{c}{\textbf{ }} & \multicolumn{1}{c}{\textbf{(ks)}} &\multicolumn{1}{c}{\textbf{(ks)}} & \multicolumn{1}{c}{\textbf{ }} & \multicolumn{1}{c}{\textbf{  }}\\ \hline 

\endfirsthead

\endfoot
\hline \hline
\endlastfoot
\hline\hline
\\
  Fornax              &    3    38   33.48 &    -35.0 29    30.5  &      100020010  &     137.5  &     68.7  &      0.004    &     CC   \\
  Antlia              &    10   30   2.21  &    -35.0 19    39.7  &     802035010  &     112.5  &     56.2  &     0.012   &     NCC  \\
  Centaurus           &    12   48   48.29 &    -41 18    47.5  &     800014010  &     61.4   &     30.7  &     0.008      &     CC    \\
  A1060               &    10   36   41.86 &    -27.0 31    51.6  &     800003010  &     64.9   &     32.4  &     0.012   &     CC   \\
  A3627               &    16   14   16.13 &    -60.0 50    59.6  &     803032010  &     87.7   &     43.8  &     0.017   &     NCC \\
  AWM7                &   02 5 29.5 	& 41 34 18 &        801035010  &     32.1   &     16.0  &     0.014    &     CC \\
  A262                &    1    52   46.13 &    36.0  9     32.8  &     802001010  &     67.6   &     33.8  &     0.017   &     CC   \\
 A3581               &    14   07    37.99 &    -27.0 01     11.6  &     807026010  &     129.4  &     64.7  &     0.022    &     CC    \\
  Coma                &    12   57   33.43 &    26.0  55    34.0  &     801097010      &     326.3 &     163.1 &     0.021    &     NCC   \\
  Ophiuchus           &    17   12   26.23 &    -23.0 22    44.4  &     802046010       &     162.8  &     81.4  &     0.029    &     CC   \\
  A2199               &    16   28   46.13 &    39.0  29    2.4   &     801056010  &     35.5   &     18.0  &     0.031    &     CC  \\
  A496		&	 04 33 38.4 & -13 15 33 .0		&	803073010 & 	68.68	& 34.4 &0.032 & CC\\
  A3571               &    13   47   26.98 &    -32.0 51    8.6   &     808094010  &     69.8   &     34.9  &     0.038    &     NCC  \\
  Triangulum Australis &    16   38   29.4  &    -64.0 20    51.7  &     803028010  &     138.8  &     69.4  &     0.048    &     NCC  \\
  A754                &    09    08    50.71 &    -9.0  38    10.0  &     802063010  &     182.8  &     91.4  &     0.054    &     NCC  \\
  A2665               &    23   50   51.86 &    6.0   08     6.7   &     801076010  &     23.1   &     11.5  &     0.099    &     NCC  \\
  A3667               &    20   12   33.84 &    -56.0 47    50.6  &     801096010  &     40.4   &     20.2  &     0.055    &     NCC  \\
  AS1101              &    23   13   59.02 &    -42.0 43    53.0  &     801093010  &     107.0  &     53.55 &     0.055     &     CC  \\
  A2256               &    17   4    3.31  &    78.0  42    40.3  &     801061010  &     188.7  &     94.3  &     0.055      &     NCC   \\
  A1831               &    13   59   12.17 &    27.0  58    9.5   &     801077010  &     32.7   &     16.3  &     0.078   &     NCC  \\
  A1795               &    13   48   53.78 &    26.0  36    3.6   &     800012010  &     19.6   &     9.8   &     0.063    &     CC \\
  A3112               &    03    17   59.57 &    -44.0 15    2.5   &     808068020  &     109.9  &     54.9  &     0.075 &     CC  \\
                 &  &      &     803054010  &     226.9  &   113.5 &      &      \\
                 &    &   &     808068010  &     108.8  &     54.4  &      &      \\
  A1800               &    13   49   26.83 &    28  05     50.3  &     801078010  &     35.6   &     17.8  &     0.075 &     NCC  \\
  A2029               &    15   10   57.82 &    05   44    59.3  &     804024010  &     13.9   &     6.9   &    0.076 &     CC  \\
  A2495               &    22   50   15.89 &    10  55    18.5  &     801080020  &     49.9   &     24.9  &     0.078        &     -    \\
  A2061               &    15   21   14.28 &    30  38    43.1  &     801081010  &     21.7   &     10.8  &     0.081 &     CC    \\
  A2249               &    17   9    50.4  &    34  29    6.4   &     801082010  &     45.6   &     22.8  &     0.085 &     -  \\
 A1750 		& 13 30 49.9 & -01 52 22 & 806095010 & 76.0 & 38.0 & 0.089 & CC \\
  A272                &    01    55   2.47  &    33  54    9.4   &     801084010  &     41.9   &     20.9  &     0.093   &     -  \\
   A2218               &    16   36   1.25  &    66  12    18.0  &     100030010  &     60.0   &     30.0  &     0.173         &     NCC \\
                 &     &   &     800019010  &     81.1   &     40.5  &              &      \\
  MS2216.0-0401       &    22   18   39.1  &    -03  46    9.5   &     807085010  &     49.7   &     24.8  &    0.094 &     -   \\
  A2142               &    15   58   7.49  &    27  17    16.4  &     801055010  &     95.0   &     47.5  &     0.091 &     NCC \\
  A2244               &    17   2    45.24 &    34  3     0.7   &     802078010  &     130.9  &     65.4  &     0.098    &     CC  \\
  A566                &    07    4    21.96 &    63  15    52.9  &     801085010  &     43.5   &     21.7  &   0.096 &     -   \\
  PKS0745-191         &    07    47   32.45 &    -19 17    24.4  &     802062010  &     55.4   &     27.7  &     0.103    &     CC   \\
  A1674               &    13   03    52.22 &    67  32    49.6  &     801062010  &     126.3  &     63.1  &     0.151    &     NCC   \\
  A2811               &    00    41   52.87 &    -28 33    18.7  &     800005010  &     53.3   &     26.6  &     0.107    &     NCC   \\
  A115                &    00    55   58.54 &    26  22    48.7  &     805077010  &     123.8  &     61.9  &     0.195    &     CC    \\
  A1246               &     11 23 50.0 & 21 25 31 &     804028010  &     88.0   &     44.0  &     0.193    &     NCC    \\
  A2219               &    16   40   17.02 &    46.0  43    12.0  &     804011010  &   170.  &     85.0  &     0.225    &     NCC   \\
  A2390		& 	21 53 35.5 & 17 41 12.0 & 804012010&  174.0 	&  87.0	& 0.228 & CC \\
  ZWCL2341.1+0000     &    23   43   38.81 &    00   19    49.1  &     803001010  &     82.2   &     41.1  &     0.273    &     NCC   \\
  A2537               &    23   8    21.48 &    -2  11   10  &     805090010  &     201.3  &     100.6 &     0.291   &     CC  \\
  Bullet              &    06    58   48.96 &    -55 55    58.8  &     801089010       &     164.8  &     82.4  &     0.297    &     NCC   \\
  A2744               &    00    14   9.53  &    -30 20    40.6  &     802033010  &     250.2  &     125.1 &     0.304    &     CC  \\
  MS1512.4+3647       &    15   14   25.42 &    36 37    11.3  &     802034010  &     466.2  &     233.1 &     0.367    &     CC   \\
  RXCJ1347.5-1145     &   13 47 30.6 & -11 45 10 &     801013010  &     114.3  &     57.1  &     0.451    &     CC  \\
    &    13   47   25.39 &    -01 10 48    34.2  &     801013020  &     138.3  &   69.1  &     0.450   & \\
\\
\hline\\
\multicolumn{8}{l}{%
  \begin{minipage}{14.cm}%
Note:  Columns are coordinates (RA, DEC), Suzaku observation ID, Exposure in front illuminated (XIS0+XIS3) and back illuminated (XIS1) observations, best-fit redshifts obtained from fits of Fe-K band of FI observations, and the category and subsample of the cluster determined based on the state of the core. NCC stands for the non-cool core sample, while CC stands for the cool-core sample. \\
  \end{minipage}%
}\\
\end{longtable*}
}
\normalsize

After the first fit iteration the $\chi^2$ improvement for the inclusion of each of these lines is determined, and the lines that are  detected with $<2\sigma$ are removed from the model. Additionally, a power-law model with an index of 1.41 and free normalization is added to the total model to account for the contribution of the cosmic X-ray background (CXB). We note that Galactic halo emission is negligible in this energy band, hence, it is not included in the model. The best-fit temperatures, normalizations of the  {\it line-free apec} models, and the fluxes of S {\sc xv}, S {\sc xvi}, Ca {\sc xix}, and Ca {\sc xx} lines are given in Table \ref{table:meas}.

{
\begin{table*}
\begin{center}
\footnotesize
\caption{Measured and Estimated Model Parameters}
\renewcommand{\arraystretch}{1.5}
\begin{tabular}{lcc|cc|cc}
\hline\hline\\
 		& Full   & 	&Cool Core &  & Non Cool-Core	   \\
		& Sample	&	&Clusters &	& Clusters  	\\
Parameters      & FI & BI   & FI   & BI & FI  & BI \\
\\\hline
\\
Measured Values:\\

kT$_{1}$  (keV)	& 5.9$\pm$0.3 & 4.8 $\pm$ 0.4 	& 6.8 $\pm$ 0.4  & 3.0 $\pm$ 0.6 & 3.1 $\pm$ 0.3    & 2.9 $\pm$ 0.2 \\

$N_{1}$ ($10^{-2}$ cm$^{-5}$)   & 1.2 $\pm$ 0.2 & 1.9 $\pm$ 0.3	& 1.5 $\pm$ 0.6   & 1.4 $\pm$ 0.7   & 1.3 $\pm$ 0.2   & 1.6 $\pm$ 0.3  \\

kT$_{2}$ (keV)	 & 8.3$\pm$0.3  & 9.56 $\pm$ 0.6  & 8.3 $\pm$ 0.6  &  9.7 $\pm$ 1.5 & 15.1 $\pm$ 1.23   & 17.1 $\pm$ 3.5  \\

$N_{2}$ ($10^{-2}$ cm$^{-5}$)   	& 1.0 $\pm$ 0.2 & 2.5 $\pm$ 0.4& 1.5 $\pm$ 0.2  & 2.9 $\pm$ 0.4   & 2.9 $\pm$ 0.7    & 2.2 $\pm$ 0.9  \\

kT$_{3}$ (keV)	 & 9.9$\pm$0.4 & --  	&  --    & -- &  -- & --  \\

$N_{3}$ ($10^{-2}$ cm$^{-5}$)   	& 1.2 $\pm$ 0.2 & --	&   --  &   --   &  -- & --  \\

S \textsc{xv} ($10^{-6}$ phts cm$^{-2}$ s$^{-1}$) & 9.9 $\pm$ 1.2  & 7.8 $\pm$ 2.0    & 16.7 $\pm$ 1.4 &  11.7 $\pm$ 4.3  & 1.4 $_{-1.4}^{+2.2}$ & 3.2 $_{-3.1}^{+4.7}$\\

S \textsc{xvi} ($10^{-6}$ phts cm$^{-2}$ s$^{-1}$) & 26.6 $\pm$ 1.1 & 24.2 $\pm$ 1.8  & 32.1 $\pm$ 1.3  & 28.7 $\pm$ 1.9  & 15.6 $\pm$ 2.5   & 16.7 $\pm$ 3.2  \\
Ar \textsc{xvii} ($10^{-6}$ phts cm$^{-2}$ s$^{-1}$) & 9.5 $\pm$ 1.1 & 7.9 $\pm$ 1.5	   &	13.2 $\pm$ 1.3   & 9.1 $\pm$ 1.6    & 5.3  $\pm$ 2.0 & 6.6 $\pm$ 2.8\\

Ca \textsc{xix} ($10^{-6}$ phts cm$^{-2}$ s$^{-1}$)& 4.6 $\pm$ 2.4  & 4.6 $\pm$ 2.0     & 7.1 $\pm$ 2.8  & 6.1 $_{-4.1}^{+3.1}$   & 4.8 $\pm$ 1.4 & 4.9 $\pm$ 2.4 \\

Ca \textsc{xx} ($10^{-6}$ phts cm$^{-2}$ s$^{-1}$)&  5.0 $\pm$ 0.7 & 6.07 $\pm$ 1.1    & 5.5 $\pm$ 0.9  & 5.5 $\pm$ 1.2   & 3.3 $\pm$ 1.2 & 3.5 $\pm$ 2.1 \\

\\
\hline
\\
Estimated Values:\\
kT (keV) 											&  3.0	& 3.3 	& 2.6 		& 2.9		& 3.8$^{*}$ 	& 3.8$^{*}$  \\
K \textsc{xviii} ($10^{-6}$ phts cm$^{-2}$ s$^{-1}$)			& 0.17 	& 0.14	& 0.21		& 0.19	& 0.12  	& 0.12 \\
Cl \textsc{xvii} ($10^{-6}$ phts cm$^{-2}$ s$^{-1}$)			& 0.08  	& 0.08  	& 0.09		& 0.09 	& 0.08	& 0.08 \\
Ar \textsc{xvii} DR ($10^{-6}$ phts cm$^{-2}$ s$^{-1}$)		& 0.03  	& 0.02  	& 0.05		& 0.03	& 0.01 	& 0.01\\

\\
\hline\hline
\\
\multicolumn{7}{l}{%
  \begin{minipage}{16.cm}%
Note: Best-fit Temperature and Normalizations of \textit{line-free apec} Model in 1.95 $-$ 6 keV fit to the Stacked XIS FI/BI spectra for various samples. The line fluxes of the S \textsc{xv}, S \textsc{xvi}, Ar \textsc{xvii}, Ca \textsc{xix}, and Ca \textsc{xx} are at the rest energies 2.51 keV, 2.63 keV, 3.12 keV, 3.90 keV, and  4.11 keV. 90\% uncertainties are given. Lower panel show the estimated maximum fluxes of the atomic lines in 3--4 keV band (before they are multiplied by a factor of 3) including K \textsc{xviii} at 3.51 keV, Cl \textsc{xvii} at 3.521 keV, and Ar \textsc{xvii} DR line at 3.62 keV. The implied plasma temperatures are calculated based on S line ratios. $^{*}$ The temperatures and line fluxes for NCC sample are determined from Ca line ratio.  \\
  \end{minipage}%
}\\
\end{tabular}
\label{table:meas}
\end{center}
\end{table*}
}

It is crucial to accurately determine the fluxes of the nearby atomic lines of K {\sc xviii}, Cl {\sc xvii}, and Ar {\sc xvii} in order to be able to measure the flux of the unidentified line at 3.5 keV. 
The line ratios of S \textsc{xv} at 2.46 keV to S \textsc{xvi} at 2.62 keV and Ca {\sc xix} at 3.9 keV to Ca {\sc xx} at 4.1 keV are good diagnostics tools for estimating plasma temperature, especially valuable for detecting the presence of cool gas (Bu14b). Following the same method presented in Bu14a, we determine the plasma temperature based on the measured fluxes of helium-like S {\sc xv} at 2.46 keV, hydrogen-like S {\sc xvi} at 2.63 keV, and helium-like Ca {\sc xix}, and hydrogen-like Ca {\sc xx} lines 3.90 keV, 4.11 keV from the spectral fits. However, the band where S {\sc xv} and S {\sc xvi} are located is crowded with strong Si {\sc xiv} lines. We therefore tie the fluxes of Si {\sc xiv} (2.01 keV: 2.37 keV: 2.51 keV) to each other with flux ratios of (21:3.5:1). The Si {\sc xiv} line ratios are estimated based on the AtomDB predictions for 3--5 keV plasma. The measured fluxes are given in the top panel of Table \ref{table:meas}.

 The maximum fluxes of the K {\sc xviii} triplet (3.47 keV: 3.49 keV: 3.51 keV) with the ratios of (1: 0.5: 2.3) are then estimated using AtomDB as described in Bu14a. Cl {\sc xvii} Lyman-$\beta$ is also included in the fits and the flux was tied to 0.15 $\times$ that of the Lyman-$\alpha$ line at 2.96 keV. We note that Cl {\sc xvii} Lyman-$\alpha$ line is not detected significantly in any of our samples, therefore, Cl {\sc xvii} Lyman-$\beta$ line is removed from our model after the first fit iteration. The maximum flux of the Ar {\sc xvii} DR line flux at 3.62 keV is determined from the measured flux of Ar {\sc xvii} triplet line at 3.12 keV. The expected flux of the Ar {\sc xvii} DR line is $<1\%$ of the Ar {\sc xvii} triplet at 3.12 keV for 3--5 keV plasma. The estimated fluxes of nearby lines (K {\sc xviii} at 3.51 keV, Cl {\sc xvii} at 3.50 keV, and  Ar {\sc xvii} DR at 3.62 keV) and plasma temperatures based on S and Ca line ratios are given in Table \ref{table:meas} bottom panel. As in Bu14a, the lower and upper limits of the fluxes of K {\sc xviii} complex, Cl {\sc xvii}, and Ar {\sc xvii} DR lines are set to 0.1 to 3 times of the maximum predicted fluxes (estimates shown in Table \ref{table:meas} are before the multiplication) to account for abundance variance between different ions.

\subsection{Full Sample}
\label{sec:fullsample}

A total of $5.06\times 10^{6}$ source counts in the 5.4 Ms FI observations and 2.9$\times 10^{6}$ source counts in the 2.7 Ms BI observations of the full sample are obtained in the 1.95 -- 6 keV energy band. The count-weighted redshift of this sample is $z\sim0.12$. After the first fit iteration with {\it line-free apec} and Gaussian models we obtain a good fit to the stacked FI observations with $\chi^{2}$ of 1032.3 for 1069 degrees-of-freedom (dof). The best-fit parameters of the model are given in Table \ref{table:meas}. The predicted plasma temperature indicated by the S \textsc{xv} to S \textsc{xv} line ratio is kT $\sim$ 3.1 keV for this sample.

To explore 3--4 keV band in the full sample (although the fit is preformed in a wider 2--6 keV band), we add an Gaussian with a fixed energy at 3.54 keV (the best-fit energy of the line detected in the {\it Suzaku} observations of the Perseus cluster). The line width is fixed to zero since we do not expect that the line width is resolved with CCD type detectors regardless of its origin. The {\it Suzaku} FI and BI detectors have energy resolution of 110--120 eV (similar energy resolution of EPIC detectors on {\it XMM-Newton}). 

Here we explore the possible interpretation of the 3.54 keV line as a decay feature of dark matter particles, therefore we use the properly weighted response files to reflect the physical properties of each cluster and the stacked sample. We note that the proper X-ray counts-weighted response files are used in modeling the continuum and known atomic transitions as described in Section \ref{sec:analysis}. 

{
\begin{table*}
\begin{center}
\caption{ Properties of Each Cluster in the Suzaku Sample}
\scriptsize
\renewcommand{\arraystretch}{1.2}
\begin{tabular}{cccccccccccc}
\hline
\\

& (1) &(2) &(3) &(4) &(5) & &(6) &(7)&(8) &(9) &(10)\\
Cluster &	 $R_{ext}$  &	 $M_{DM}^{proj}$ &	 $M_{DM}^{proj}/D_{L}^{2}$ &	 $\omega_{cnt}$ &	 $\omega_{DM}$ 	& Cluster &	 $R_{ext}$  &	 $M_{DM}^{proj}$ &	 $M_{DM}^{proj}/D_{L}^{2}$ &	 $\omega_{cnt}$ &	 $\omega_{DM}$ \\		
 &	 (Mpc)  &	 ($10^{14}M_{\odot}$) &	 ($\frac{10^{10}M_{\odot}}{Mpc^{2}}$) &($10^{-2}$)& ($10^{-2}$)		& &	 (Mpc)  &	 ($10^{14}M_{\odot}$) &	($\frac{10^{10}M_{\odot}}{Mpc^{2}}$) &($10^{-2}$)	& ($10^{-2}$)\\			
\\													
  \hline
  \\													
Fornax &	 0.04 &	 0.04 &	 1.18 &	 1.34 &	4.97	&	A2029 &	 0.71 &	 7.17 &	 0.62 &	 0.40 &	 0.09 \\ 	
Antlia &	 0.10 &	 0.18 &	 0.98 &	 0.61 &	3.41	&	A2495 &	 0.69 &	 2.63 &	 0.24 &	 0.29 &	 1.02 \\ 	
Centaurus &	 0.12 &	 0.41 &	 1.55 &	 3.18 &	 2.93	&	A2061 &	 0.72 &	 3.89 &	 0.33 &	 0.15 &	 0.17 \\ 	
A1060 &	 0.13 &	 0.39 &	 1.19 &	 1.66 &	2.38	&	A2249 &	 0.74 &	 5.64 &	 0.45 &	 0.28 &	  0.49 \\ 	
A3627 &	 0.16 &	 1.03 &	 2.33 &	 1.87 &	6.34	&	A1750 &	 0.78 &	 2.34 &	 0.16 &	 0.29 &	 1.12 \\ 	
AWM7 &	 0.16 &	 0.57 &	 1.20 &	 1.27 &	1.19	&	A272 &	 0.78 &	 2.58 &	 0.18 &	 0.20 &	 0.22 \\ 	
A262 &	 0.16 &	 0.37 &	 0.79 &	 1.04 &	1.65	&	MS2216.0-0401 &	 0.80 &	 2.35 &	 0.15 &	 0.23&	 0.29 \\ 	
A3581 &	 0.23 &	 0.48 &	 0.49 &	 1.33 &	1.96	&	A2142 &	 0.82 &	 9.11 &	 0.57 &	 2.85 &	 0.48 \\ 	
Coma &	 0.23 &	 1.81 &	 1.81 &	 17.71 &	18.42	&	A2244 &	 0.86 &	 5.65 &	 0.31 &	 1.64 &	 2.48 \\ 	
Ophiuchus &	 0.27 &	 2.76 &	 1.99 &	 20.39 &	10.18	&	A566 &	 0.87 &	 3.32 &	 0.18 &	 0.16 &	 0.45 \\ 	
A2199 &	 0.29 &	 2.03 &	 1.26 &	 1.32 &	1.40	&	PKS0745-191 &	 0.91 &	 7.76 &	 0.37 &	 1.57 &	 0.33 \\ 	
A496 &	 0.31 &	 1.34 &	 0.71 &	 2.22 &	1.52	&	A1674 &	 0.91 &	 3.14 &	 0.14 &	 0.18 &	 1.63 \\ 	
A3571 &	 0.38 &	 2.79 &	 0.98 &	 3.19 &	1.56	&	A2811 &	 0.94 &	 4.83 &	 0.21 &	 0.21 &	 0.25 \\ 	
Triangulum  &	 0.48 &	 5.38 &	 1.12 &	 5.98 &	4.33	&	A2218 &	 1.20 &	 7.63 &	 0.11 &	 0.46&	 1.07 \\ 	
 Australis &		&&&&		&				A1246 &	 1.15 &	 6.86 &	 0.09 &	 0.33 &	  0.36 \\ 	
A754 &	 0.52 &	 5.68 &	 1.01 &	 4.66 &	6.59	&	A115 &	 1.50 &	 15.40 &	 0.18 &	 0.94 &	 0.38 \\ 	

A3667 &	 0.52 &	 3.65 &	 0.65 &	 0.93 &	0.78	&	A2390 &	 1.38 &	 12.34 &	 0.10 &	 1.43 &	 1.15 \\ 	
A2665 &	 0.51 &	 4.30 &	 0.79 &	 0.15 &	0.84	&	A2219 &	 1.65 &	 21.20 &	 0.18 &	 1.44 &	  0.67 \\ 	
AS1101 &	 0.53 &	 1.45 &	 0.24 &	 1.50 &	2.73	&	ZWCL2341.1+0000 &	 1.07 &	 6.08 &	 0.03 &	 0.07 &	 0.57 \\ 	
A2256 &	 0.54 &	 4.98 &	 0.80 &	 4.00 &     1.46	&	A2537 &	 1.32 &	 11.73 &	 0.05 &	 0.34 &	  0.27 \\ 	
A1831 &	 0.58 &	 1.87 &	 0.26 &	 0.23 &	0.86	&	Bullet &	 1.58 &	 20.15 &	 0.09 &	 0.98 &	 0.35 \\ 	
A1795 &	 0.58 &	 4.07 &	 0.55 &	 0.80 &	 0.16	&	A2744 &	 1.30 &	 11.36 &	 0.05 &	 0.82 &	 0.91 \\ 	
A3112 &	 0.68 &	 3.42 &	 0.32 &	 7.54 &	7.97	&	MS1512.4+3647 &	 0.76 &	 2.46 &	 0.01 &	 0.20 &	 0.93 \\ 	
A1800 &	 0.69 &	 2.41 &	 0.22 &	 0.22 &	0.37	&	RXCJ1347.5-1145 &	 1.57 &	 23.34 &	 0.04 &	 1.17 &	 0.07 \\ 	
\\
\hline\hline
\\
\multicolumn{12}{l}{%
  \begin{minipage}{18.cm}%
Note:  Columns (1) and (6) show the spectral extraction radius in Mpc, Columns (2) and (7) are the estimated projected dark matter masses in the spectral extraction radii $M_{DM}^{proj} (R_{ext})$, the projected dark matter mass per luminosity distance $M_{DM}^{proj}/D_{L}^{2}$ are given in columns (3) and (8), columns (4), (5), (9), and (10) show the weighting factors ($\omega_{cnt}$) calculated based on the total counts in the fitting band 2--6 keV and the weighting factors ($\omega_{dm}$) calculated based on the predicted dark matter flux. These factors are used in the stacking of ARFs and RMFs of each cluster in the sample. \\
  \end{minipage}%
}\\
\end{tabular}
\label{table:prop}
\end{center}

\end{table*}
}

The contribution of each cluster to any flux due to dark matter decay in the stacked sample is related to
the mass of decaying dark matter particles within the FOV.
Following the same formulation laid out by Bu14a, the weight of each cluster in the full {\it Suzaku} sample
 is;
\begin{equation}
\omega_{i,dm} = \frac{M_{i, DM}^{proj}(<R_{ext})(1+z_{i})}{4\pi D_{i,L}^{2}}\ \frac{e_{i}}{e_{tot}},
\label{eqn:weight}
\end{equation}

\noindent where $z_{i}$ is the redshift of the \textit{i}th cluster, and $e_{i}$ and $e_{tot}$
are the exposure time of the \textit{i}th cluster and the total exposure time of
the sample, $M_{DM}^{FOV}$ is the
projected dark matter mass within the spectral extraction region ($R_{ext}$,
which is either $R_{500}$ or $R_{FOV}$), and $D_{L}$ is the luminosity distance. We use
the  the Navarro--Frenk--White (NFW) profile \citep{navarro1997} to determine the dark matter mass
within the field-of-view. The steps in these calculation are described in detail in B14a. The calculated weight of each cluster is given in Table \ref{table:prop} for each cluster in the full {\it Suzaku} sample.

Initially, we examine the 3--4 keV band of the stacked FI observations of the full sample. After the addition of the Gaussian model at 3.54 keV, the new best-fit $\chi^{2}$ becomes 1028.1 for 1068 dof. The change in the $\chi^{2}$ is 4.1 after the addition of a degree of freedom. The best-fit flux of the line is  1.0$_{-0.5}^{+0.5}\ (_{-0.9}^{+1.3}) \ \times\ 10^{-6}$ phts cm$^{-2}$ s$^{-1}$.  The change in the $\chi^{2}$ corresponds to a 2$\sigma$ detection for an additional degree of freedom in the stacked FI observations of the full sample. The stacked XIS FI spectrum of the full sample and the best-fit models before and after the Gaussian line is added are shown in Figure \ref{fig:spec} left panel.

For the BI observations of the full sample, the fit with {\it line-free apec} model and additional Gaussians for known atomic lines give a good-fit with $\chi^{2}$ of 1111.5 (1078 dof). The line is not detected at a statistically significant level in this spectrum. Additional Gaussian line at 3.54 keV improves the fit by $\Delta \chi^{2}$= 1.5 for an extra dof  (the $\chi^{2}$ becomes 1109.9 for 1077 dof). The best-fit flux of the line is 9.1$^{+1.5}_{-7.3}\ (^{+2.2}_{-9.1})\ \times\ 10^{-7}$ phts cm$^{-2}$ s$^{-1}$. The stacked XIS BI spectrum of the full sample and the best-fit models before and after the Gaussian line is added at 3.54 keV are shown in Figure \ref{fig:spec} right panel.

To test the decaying dark matter origin of the signal, we further investigate if the mixing angles indicated by these fluxes are consistent with the previous detections in the literature. The measured flux from a mass of dark matter within the FOV can be converted into the decay rate assuming dark matter particles decaying monochromatically with E$_{\gamma}$= m$_{s}$/2. The mixing angle for this decay is 

\begin{equation}
\begin{split}
\rm {sin}^{2} (2\theta)= \frac{F_{DM}}{12.76\ \rm{cm^{-2}} \ s^{-1}}
\left( \frac{10^{14}\ M_{\odot}}{M_{DM}^{FOV}}\right) \\
\left( \frac{D_{L}}{100\ \rm{Mpc}}\right)^{2} \left( \frac{1}{1+z}\right)
\left( \frac{1\ \rm{keV}}{m_{s}}\right)^{4},
\end{split}
\label{eqn:mixangle}
\end{equation}
\noindent where $F_{DM}$ is the observed flux due to dark matter decay \citep{pal82} and is related to
the surface density or flux of decaying dark matter particles within the FOV;

\begin{equation}
F_{DM} = \frac{M_{DM}^{FOV} }{4\pi D_{L}^{2} } \frac{\Gamma_{\gamma}}{m_{s}} (1+z)\ \  \ \rm{photons\ cm^{-2}\ s^{-1}},
\label{eqn:dmflux}
\end{equation}
\noindent where $\Gamma_{\gamma}$ and $m_{s}$ are the decay rate and dark matter particle mass, respectively.

{
\begin{table*}[]
\begin{center}
\caption{Measured Flux of the 3.5 keV Line in the Stacked  {\it Suzaku} Clusters}
\renewcommand{\arraystretch}{1.5}
\begin{tabular}{ccccccccc}
\hline\hline\\
			& (1) &(2) &(3) &(4) &(5) &(6) &(7)\\
Sample			& Inst.  & 	Energy	& Flux	& $\chi^{2}$	& $\Delta \chi^{2}$  &  $M_{DM}^{proj}/D_{L}^{2}$ & $\sin^{2}(2\theta)$\\
				& 		& (keV)	& ($10^{-6}$ phts cm$^{-2}$ s$^{-1}$)	&	(dof)	& (dof) & ($10^{10}$ M$_\odot/Mpc^{2}$) & (10$^{-11}$)\\			
\\\hline
\\
Full Sample	& FI	& 3.54	& 1.0$_{-0.5}^{+0.5}\ (_{-0.9}^{+1.3})$ & 1028.1 (1068)	& 4.11 (1)	 & 1.17 	& 2.7$_{-1.4}^{+1.4}\ (_{-2.3}^{+3.4})$ \\
			& BI	& 3.54 	& 0.9$^{+0.2}_{-0.7}\ (^{+0.2}_{-0.9})$ & 1109.9 (1077) 	& 1.46 (1)	& 1.17	& 2.5$_{-1.9}^{+0.4}\ (_{-2.5}^{+5.9})$\\
\\
\hline
\\
Cool-Core 	& FI	& 3.54 	& $<$1.4 	& 1131.7 (1069) & 1.68 (1) & 1.06	& $<$5.1 \\
Clusters 		& BI	& 3.54 	& $<$2.1		& 1143.0 (1072) & 0.15 (1) & 1.06 	&  $<$6.1\\
\\
\hline
\\
Non-Cool Core	& FI 	& 3.54 	& 2.0$_{-0.7}^{+1.0}$ ($_{-1.2}^{+1.9}$) & 1034.7 (1075) & 6.56  (1)	& 1.19	& 5.3$_{-1.8}^{+2.6}\ (_{-3.1}^{+4.7})$ \\
Clusters		& BI 	& 3.54	& $<$5.4   & 1159.9 (1072) & 0.51 (1)	& 1.19	& $<$14.1\\
	
\\
\hline\hline
\\
\multicolumn{8}{l}{%
  \begin{minipage}{16.cm}%
Note: Columns (2) and (3) are the rest energy and flux of the unidentified line in the units of photons cm$^{-2}$ s$^{-1}$ at the 68\% (90\%) confidence level. Column (4) and (5) show the $\chi^{2}$ after the line is added to
the total model and change in the  $\chi^{2}$ when an additional Gaussian component is added to the fit; column (6) is the weighted ratio of mass to distance squared of the samples, and column (7) shows the mixing angle limits measured in each sample. Reported constraining limits are at 90\% confidence. Energies are held fixed during the model fitting. \\
  \end{minipage}%
}\\
\end{tabular}
\label{table:results}
\end{center}
\end{table*}
}

Using $\omega_{dm}$ and the projected dark matter masses given in Table \ref{table:prop}, we find that the weighted projected
dark matter mass per distance squared of the full {\it Suzaku} sample is $1.17\times 10^{10}$ M$_\odot/Mpc^{2}$. Using Equation \ref{eqn:mixangle}, one can calculate the mixing angle to be $\sin^{2}(2\theta)= 2.7_{-1.4}^{+1.4}\ (_{-2.3}^{+3.4})\times 10^{-11}$ for the full {\it Suzaku} FI sample for a particle mass of $m_s$ = 7.08 keV. The associated 90\% upper limit to the mixing angle is $\sin^{2}(2\theta)<6.1\times 10^{-11}$ in this sample. 

To compare the consistency between XIS FI spectrum and the previously detected line flux in {\it XMM-Newton} observations (Bu14a), we scale the flux based on the signal from the larger cluster sample under the dark matter decay scenario. Figure \ref{fig:moslimit} shows the zoomed in 3.3--3.8 keV band of the stacked XIS full FI sample spectrum. The solid line marks the best-fit flux of the 3.54 keV line scaled from the Bu14a full sample flux with the 90\% uncertainties are marked with dashed lines.  As the figure clearly shows the XIS FI observations are consistent with the {\it XMM-Newton} observations at a 90\% level. 

The {\it Suzaku} BI observations of the full sample give a mixing angle measurement of $\sin^{2}(2\theta)=2.5_{-1.9}^{+0.4}\ (_{-2.4}^{+5.9}) \times 10^{-11}$ for the same weighted mass per distance squared. These are given in Table \ref{table:results}. The {\it Suzaku} full FI/BI sample measurements are consistent with each other. The mixing angle measured from the full {\it XMM-Newton} MOS/PN samples ($\sin^{2}(2\theta)=6.8^{+1.4}_{-1.4}\ (_{-3.0}^{+2.0})\times 10^{-11}$) are consistent at a 1$\sigma$ confidence level and the MOS observations of bright clusters (Coma+Ophiuchus+Centaurus; $\sin^{2}(2\theta)=1.8^{+0.44}_{-0.39}\ (_{-1.2}^{+1.2}) \times 10^{-10}$) are consistent at a $\sim2.7\sigma$ confidence level (see Bu14a). The core excised observations of the Perseus cluster ($2.3^{+0.7}_{-0.7}\ (_{-1.2}^{+1.2}) \times 10^{-10}$; see Bu14a) measurement is in tension with the present {\it Suzaku} sample result at a level of $\sim2.5\sigma$. Comparison of mixing angles measured from {\it Suzaku} samples with the previous detections and limits are shown in Figure \ref{fig:limits}.

 \begin{figure*}
 \centering
\hspace{-4mm}\includegraphics[width=18.cm, angle=0]{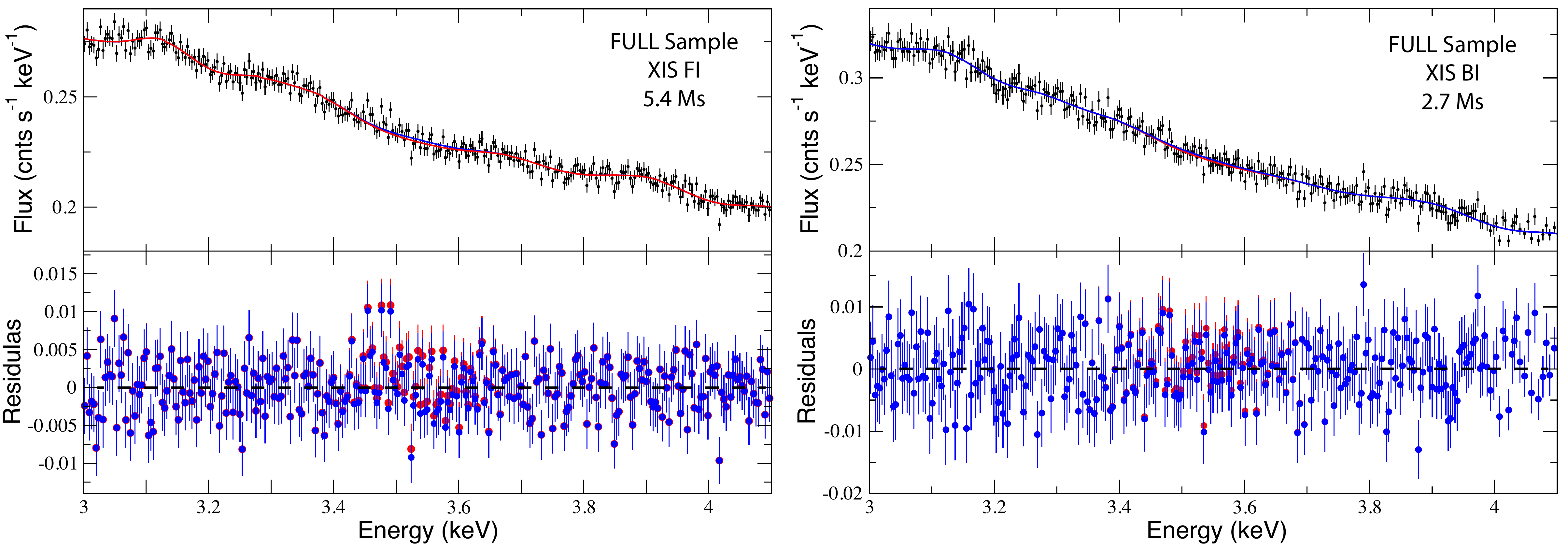}
\caption{3--4 keV band of the binned stacked {\it Suzaku} XIS FI (left panel) and XIS BI (right panel) spectra of the full sample. The figures show the energy band where the unidentified 3.5 keV line is detected by Bu14a. The Gaussian lines with maximum values of the flux normalizations of K {\sc xviii} and Ar {\sc xvii} DR are already included in the models. The 3.5 keV line is not significantly detected in either of these samples. The red and blue model lines in the top panels show the total model before and after a Gaussian line is added at 3.54 keV. Bottom panels show the residuals before (red) and after (blue) the Gaussian line is added.}
\label{fig:spec}
\end{figure*}

 \subsection{Cool-Core Clusters}
 
We now divide the full sample into two independent subsamples in order to investigate if the line flux correlates with the presence cool gas in the intra-cluster plasma. The clusters are divided into cool-core clusters (CC) and non cool-core clusters (NCC) based on previous identifications in the literature. If, indeed, the flux of the 3.5 keV line is stronger in the stacked cool-core cluster sample (i.e. if a correlation is observed between gas temperature and the flux), this would be a strong indication that the 3.5 keV line is astrophysical in origin. The classification of each cluster is given in Table \ref{table:obs}. For some of the clusters in the full sample (e.g., A2495, A2249, A272, RXC J2218.8-0258, MS 2216.0-0401, and A566) the X-ray studies with high angular resolution observatories, e.g. {\it Chandra} and {\it XMM-Newton}, are not available in the literature. Due to relatively large point-spread-function ($\sim2^\prime$ half-power diameter) of {\it Suzaku} mirrors, we cannot distinguish if these clusters have cool intra-cluster gas in their center. Hence we exclude these clusters from both subsamples.

We have performed the stacking process following the same approach outlined in Section \ref{sec:analysis} for the CC clusters.
A total of 3.1 Ms of good stacked FI and 1.5 Ms BI observations are obtained in this subsample. The weighted mean redshift of the subsample is 0.13. The stacked FI/BI observations of this subsample contain 52\% and 51\% of the total source counts of the full FI and BI observations.

We fit the stacked {\it Suzaku} FI spectra of the CC cluster as described in Section \ref{sec:results}. The best-fit temperatures, normalizations and the fluxes of S {\sc xv}, S {\sc xvi}, Ca {\sc xix}, and Ca {\sc xx} are given in Table \ref{table:meas}.
Cl Ly-$\alpha$ at 2.96 keV is not detected significantly in this spectrum, we therefore exclude Cl Ly-$\beta$ line at 3.51 keV in our fits. Overall we obtain a good-fit to the stacked CC spectrum with $\chi^{2}=$1130.0 (1068 dof).
Adding in an extra Gaussian model to the MOS spectrum at 3.54 keV does not improve the fit significantly ($\Delta\chi^{2}=1.68$) for an additional degrees-of-freedom and results in a non-detection. The 90\% upper limit on the flux of this line at 3.54 keV is $1.4\times 10^{-6}$ photons cm$^{-2}$ s$^{-1}$ from this spectrum. The upper limit on the flux can be translated to a mixing angle of $5.1\times 10^{-11}$  for a given projected dark matter mass per distance squared for the sample (1.06$\times 10^{10}$ M$_{sun}$/Mpc$^{2}$). The mixing angle indicated by the stacked  FI observations of CC clusters 
is consistent with the {\it Suzaku} full-sample and the previous {\it XMM-Newton} detections.

We note that the discrepancy observed in plasma temperatures between FI and BI observations of the cool-core clusters might be due to the difference in the response of the FI and BI sensors, or the power-law normalizations for CXB which were left free during the fits. We note that spectra of individual clusters are rescaled to their emitter frame before being stacked. Therefore, the stacked spectra do not contain any physical meaning after the blue-shifting and stacking processes. The main goal of this work is to model the continuum accurately to make the analysis sensitive to faint line detections. Therefore, the observed difference is not worrying in the context of this work. The crucial point is that the line ratios observed in FI and BI observations within each sample are consistent. The line ratios are used to determine the plasma temperature and fluxes faint lines in 2.5--4.1 keV band.

The overall fit to the stacked BI observations to CC clusters is acceptable with $\chi^{2}$ of 1142.85 for 1068 dof. Adding an extra Gaussian line at 3.54 keV 
does not improve the fit significantly and results in a non-detection. The 90\% upper limit to the flux is $2.1\times 10^{-6}$ photons cm$^{-2}$ s$^{-1}$  from this spectrum; the upper limit on the mixing angle ($<6.1\times 10^{-11}$M$_{sun}$/Mpc$^{2}$) from this flux limit is consistent with the full-sample and FI detections.

 \begin{figure}
 \centering
\hspace{-4mm}\includegraphics[width=9.cm, angle=0]{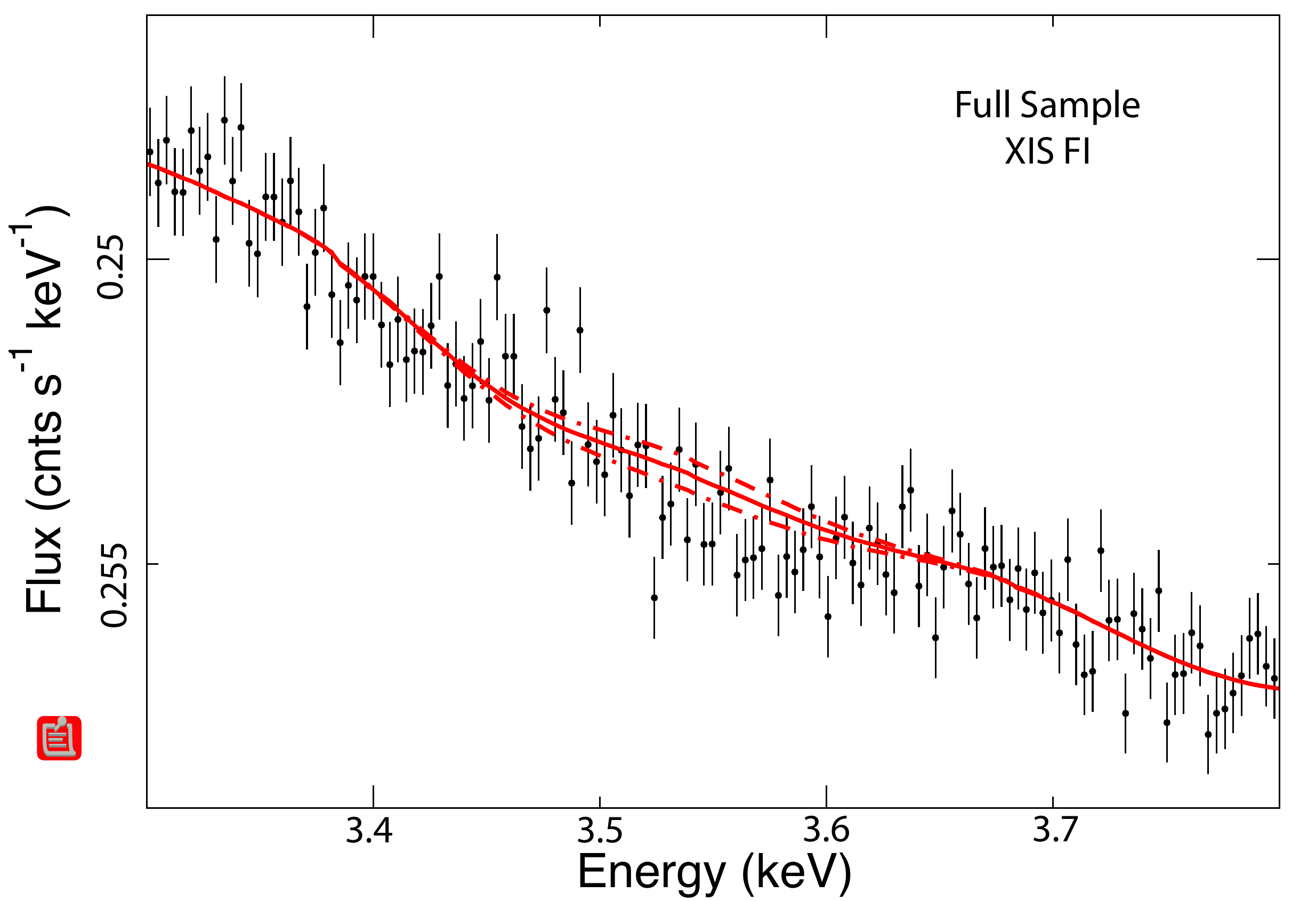}
\caption{ Zoomed in {\it Suzaku} XIS FI spectrum of the stacked full sample. The solid red line shows the best-fit  line flux scaled from the {\it XMM-Newton} MOS full sample detection ($4\times10^{-6}$ phot s$^{-1}$ cm$^{-2}$) under the dark matter decay scenario. The dashed lines mark the 90\% confidence levels of the scaled flux.}
\label{fig:moslimit}
\end{figure}

\subsection{Non-Cool Core Clusters}

We now examine the  FI and BI observations of the NCC clusters. A total of 2.2 Ms good FI and 1.1 Ms good BI observations are obtained for this sample.
The NCC cluster sample contain 46\% of the total FI source counts, 45\% of the total BI source counts of the full sample. 
The redshift has a weighted mean value at 0.11 and the projected dark matter mass per distance squared is $1.19\times10^{10}$ M$_{sun}$/Mpc$^{2}$ of the NCC subsample.

To be able to estimate the fluxes of K {\sc xviii}, Cl {\sc xvii}, and Ar {\sc xvii} lines conservatively, we use Ca {\sc xix} and Ca {\sc xx} lines for this sample. Probing the 3--4 keV band the FI observations does not reveal significant residuals around 3.54 keV.
Indeed, the first fitting attempt (without an Gaussian model at 3.54 keV) is an overall good with $\chi^{2}$ of 1041.3 for 1076 dof. Addition of a Gaussian model improves the fit by  $\Delta \chi^{2}$ of 6.56 for an extra dof. The best-fit flux of the line becomes 2.0$_{-0.7}^{+1.0}$ ($_{-1.2}^{+1.9}$) $\times10^{-6}$~photons cm$^{-2}$ s$^{-1}$. The mixing angle corresponding to this flux is 5.3$_{-1.8}^{+2.6}\ (_{-3.1}^{+4.7})\times 10^{-11}$, which is consistent with the full sample. The 90\% upper limit to its flux is $3.9\times10^{-6}$ phts~cm~$^{-2}$~s$^{-1}$ in the FI observations of the non-cool clusters with a mixing angle of $1.0\times 10^{-10}$.

Treating the stacked BI observations of the NCC clusters, we obtain an acceptable fit ($\chi^{2}$of 1159.4 with 1071 dof) without an additional Gaussian model at 3.54 keV. Adding an extra Gaussian component at 3.54 keV changes the goodness of the fit by $\Delta\chi^{2}$ of 0.51 ($\Delta dof=1$). The 90\% upper limits to the flux of the line is $5.4\times10^{-6}$~phts cm$^{-2}$ s$^{-1}$, which corresponds to a mixing angle of $1.4\times 10^{-10}$ for this sample.

 \begin{figure}
 \centering
\hspace{-4mm}\includegraphics[width=9.cm, angle=0]{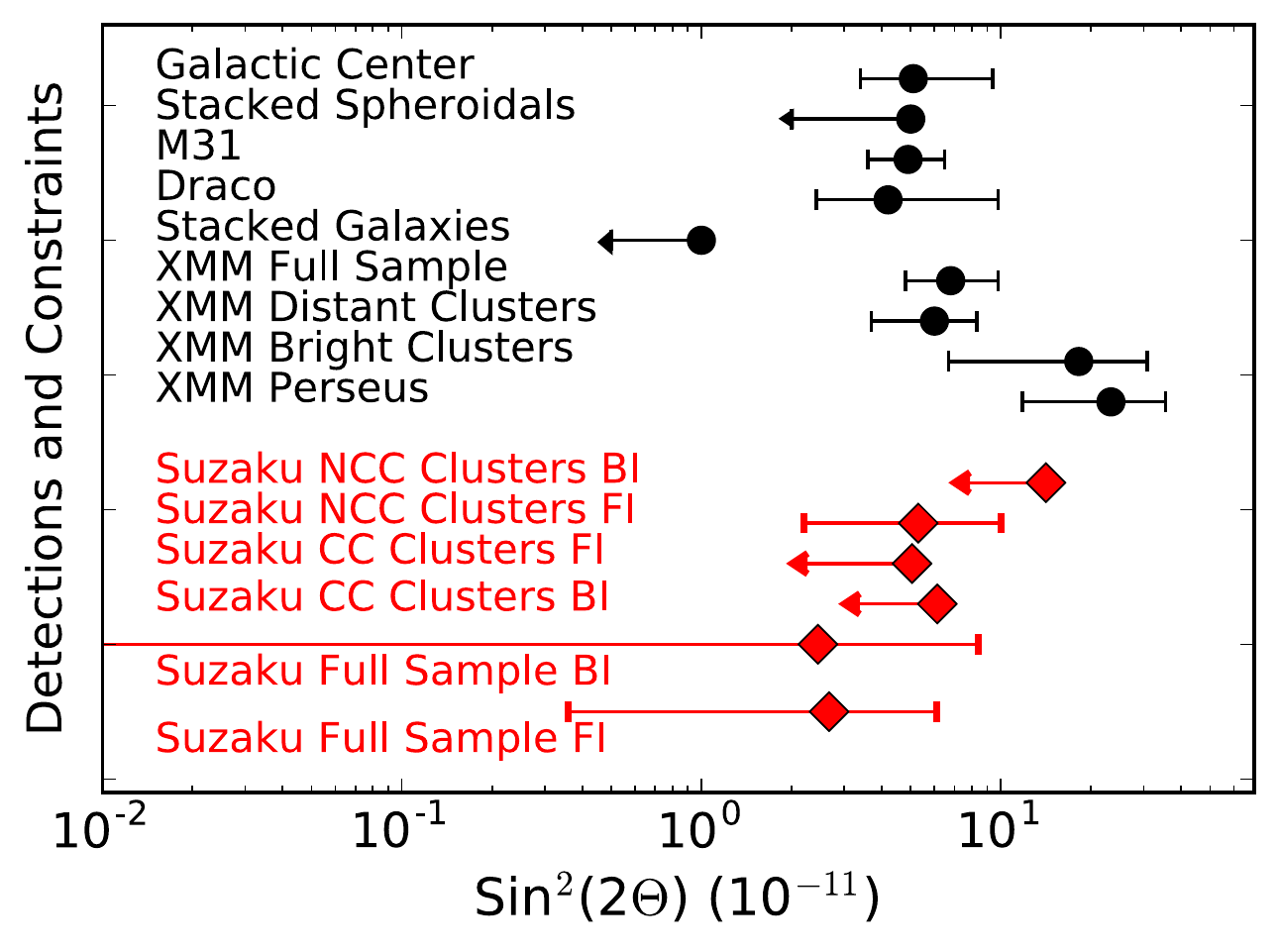}
\caption{A comparison of sterile neutrino mixing angle upper limits obtained from the stacked galaxy clusters observed with {\it Suzaku}. The results in the literature are also shown. The error bars and upper limits from this work and Bu14a results are 90\% confidence levels. The upper limits from the stacked Spheroidal galaxies \citep[][2$\sigma$]{malyshev14} and stacked galaxies \citep[][90\%]{anderson15} together with the detections in the Galactic Center \citep[][90\%]{bo15}, the Draco dwarf spheroidal \citep[][1$\sigma$]{ru15}, M31 (Bo14a, 1$\sigma$)
are shown. Anomalously high Perseus flux reported in Bu14a is clearly seen in the Figure. We note that the particle mass is not compared here.}
\label{fig:limits}
\end{figure}

\section{Summary}
\label{sec:intro}

Stacking X-ray spectra of galaxy clusters at different redshifts provides a sensitive tool to detect weak emission features. This method, tested on the {\it XMM-Newton} observations of 73 clusters (Bu14a), resulted in the detection of a very weak unidentified spectral line at $\sim3.5$~keV. In this work, we take a similar approach and stack {\it Suzaku} FI (XIS0, XIS3) and BI (XIS1) observations of 47 nearby ($0.01<z<$0.45) galaxy clusters to look for the unidentified emission line. Our {\it Suzaku} sample consists of 5.4 Ms of FI and 2.1 Ms of BI observations. The total source counts collected in this study is less than those of the stacked {\it XMM-Newton} observations by a factor of 1.8. The redshift span is slightly larger of the {\it Suzaku} full sample than the full {\it XMM-Newton} sample, leading to more effective smearing of the instrumental features. The redshift range of the {\it Suzaku} full sample corresponds to an energy difference of up to 1.44 keV at 3.5 keV, which is sufficient to smear out and eliminate the background or response features.

The stacked FI data for the full sample prefers an additional emission line at $E=3.54$ keV (the energy fixed at the best-fit value for the {\it Suzaku} line detection in Perseus Franse et al.\ (2016)), but only
at 2$\sigma$ confidence level with a flux of $1.0_{-0.5}^{+0.5}\ (_{-0.9}^{+1.3}) \ \times\ 10^{-6}$~phts~cm$^{-2}$~s$^{-1}$. The statistics of the dataset is insufficient to constrain the energy of this faint line. The line is not significantly detected in the BI observations, however an additional Gaussian model improves the fit by $\Delta\chi^{2}=1.5$ and has a flux of $9.1^{+1.5}_{-7.3}\ (^{+2.2}_{-9.1}) \ \times\ 10^{-6}$~phts~cm$^{-2}$~s$^{-1}$. The fluxes observed in FI and BI observations are in agreement with each other.

In an attempt to investigate a possible correlation of the flux of the unidentified line with cooler gas in the ICM, we divide the full sample into two subsamples; CC and NCC clusters. If a correlation is observed, it would be an indication that the unidentified line is astrophysical in origin. Atomic lines are more prominent in cool-core clusters where a significant amount of cooler gas with higher metal abundances resides in the core. However, we do not detect any significant spectral feature at 3.5 keV in the separate CC and NCC clusters. The FI observations of the NCC sample shows a  weak 2.4$\sigma$ residual at 3.54 keV, with a flux of 5.3$_{-1.8}^{+2.6}\ (_{-3.1}^{+4.7})\times10^{-6}$~phts~cm$^{-2}$~s$^{-1}$. The upper limits derived from these samples are consistent with previous detections. We note that both CC and NCC subsamples contain fewer number of source counts compared to all of the {\it XMM-Newton} samples studied in Bu14a so the sensitivity of the presented {\it Suzaku} analysis is weaker. We also note that due to smaller FOV and lower effective area of  the {\it Suzaku} XIS detectors compared to the {\it XMM-Newton} EPIC detectors, this analysis might be less sensitive to a weak signal from dark matter decay. The value of this analysis is in that it is independent and performed with a different instrument. 

The upper limits provided by this work (full sample; $\sin^{2}(2\theta)=6.1\times 10^{-11}$) is in agreement with the detections in the combined M31, Galactic center observations ($\sin^{2}(2\theta)=5-7\times 10^{-11}$; see Boyarsky et al.\ 2015), and results from deep MOS ($\sin^{2}(2\theta)<5.8\times10^{-11}$) and PN  ($\sin^{2}(2\theta)=1.8-8\times10^{-11}$) observations of the Draco galaxy \citep{ru15}. However, the line flux in the core of the Perseus cluster is in tension with the presented stacked {\it Suzaku} and {\it XMM-Newton} clusters and other detections \citep[Bu14a, ][]{franse16}. Studying the origin of the 3.5 keV line with CCD resolution observations of galaxy clusters and other astronomical objects appears to have reached its limit; the problem requires higher-resolution spectroscopy such as that expected from {\it Hitomi} (Astro-H).

%

\section{Acknowledgements}
Authors thank Keith Arnaud for providing help with response remapping and the anonymous referee for useful comments on the draft. Support for this work was provided by NASA through contract NNX14AF78G, NNX13AE77G, and NNX15AC76G. E. Miller, and M. Bautz acknowledge support from NASA grants NNX13AE77G and NNX15AC76G. A. Foster acknowledges NASA grant NNX15AE16G. Support for SWR was provided by the Chandra X-ray Center through NASA contract NAS8-03060 and the Smithsonian Institution.


\begin{thebibliography}{34}
\expandafter\ifx\csname natexlab\endcsname\relax\def\natexlab#1{#1}\fi

\bibitem[Abazajian(2014)]{abazajian14} Abazajian, K.~N.\ 2014, Physical Review Letters, 112, 161303 

\bibitem[Anderson et al.(2015)]{anderson15} Anderson, M.~E., Churazov, E., \& Bregman, J.~N.\ 2015, \mnras, 452, 3905 

\bibitem[Arnaud(1996)]{arnaud96} Arnaud, K.~A.\ 1996, 
Astronomical Data Analysis Software and Systems V, 101, 17 

\bibitem[Boyarsky et al.(2014)]{bo14} Boyarsky, A., 
Ruchayskiy, O., Iakubovskyi, D., \& Franse, J.\ 2014, Physical Review Letters, 113, 251301 

\bibitem[Boyarsky et al.(2015)]{bo15} Boyarsky, A., Franse, 
J., Iakubovskyi, D., \& Ruchayskiy, O.\ 2015, Physical Review Letters, 115, 161301 

\bibitem[Bulbul et al.(2014a)]{b14} Bulbul, E., Markevitch, 
M., Foster, A., et al.\ 2014, \apj, 789, 13 

\bibitem[Bulbul et al.(2014b)]{b14b} Bulbul, E., Markevitch, 
M., Foster, A.~R., et al.\ 2014, arXiv:1409.4143 

\bibitem[Bulbul et al.(2016)]{bu2016} Bulbul, E., Randall, 
S.~W., Bayliss, M., et al.\ 2016, \apj, 818, 131 

\bibitem[Carlson et al.(2015)]{carlson15} Carlson, E., Jeltema, T., \& Profumo, S.\ 2015, JCAP, 2, 009 

\bibitem[Franse et al.(2016)]{franse16} Franse, J., Bulbul, E., Foster, A., et al.\ 2016, arXiv:1604.01759 

\bibitem[Foster et al.(2012)]{foster12} Foster, A.~R., Ji, L., 
Smith, R.~K., \& Brickhouse, N.~S.\ 2012, \apj, 756, 128 

\bibitem[Horiuchi et al.(2016)]{horiuchi16} Horiuchi, S., Bozek, B., Abazajian, K.~N., et al.\ 2016, \mnras, 456, 4346 

\bibitem[Iakubovskyi et al.(2015)]{iakubovskyi15} Iakubovskyi, D., Bulbul, E., Foster, A.~R., Savchenko, D., \& Sadova, V.\ 2015, arXiv:1508.05186 

\bibitem[Jeltema 
\& Profumo(2015)]{jp15} Jeltema, T., \& Profumo, S.\ 2015, \mnras, 450, 2143 

\bibitem[Gu et al.(2015)]{gu15} Gu, L., Kaastra, J., Raassen, A.~J.~J., et al.\ 2015, \aap, 584, L11 

\bibitem[Malyshev et al.(2014)]{malyshev14} Malyshev, D., Neronov, A., \& Eckert, D.\ 2014, \prd, 90, 103506 


\bibitem[Navarro et al.(1997)]{navarro1997} Navarro, J.~F., Frenk, 
C.~S., \& White, S.~D.~M.\ 1997, \apj, 490, 493 

\bibitem[Smith et al.(2001)]{smith01} Smith, R.~K., Brickhouse, 
N.~S., Liedahl, D.~A., \& Raymond, J.~C.\ 2001, \apjl, 556, L91 

\bibitem[Pal \& Wolfenstein(1982)]{pal82} Pal, P.~B., \& Wolfenstein, L.\ 1982, \prd, 25, 766 


\bibitem[Protassov et al.(2002)]{protassov02} Protassov, R., van Dyk, D.~A., Connors, A., Kashyap, V.~L., \& Siemiginowska, A.\ 2002, \apj, 571, 545 

\bibitem[Ruchayskiy et al.(2015)]{ru15} Ruchayskiy, O., 
Boyarsky, A., Iakubovskyi, D., et al.\ 2015, arXiv:1512.07217 

\bibitem[Tamura et al.(2015)]{tamura15} Tamura, T., Iizuka, R., Maeda, Y., Mitsuda, K., \& Yamasaki, N.~Y.\ 2015, \pasj, 67, 23 


\bibitem[Urban et al.(2015)]{urban15} Urban, O., Werner, N., 
Allen, S.~W., et al.\ 2015, \mnras, 451, 2447 
\bibitem[Vikhlinin et al.(2009)]{vikhlinin2009} Vikhlinin, A., 
Burenin, R.~A., Ebeling, H., et al.\ 2009, \apj, 692, 1033 

\end{thebibliography}
\end{document}